\documentclass[useAMS,usenatbib]{mn2e}
\usepackage{epsfig}
\usepackage{rotating}
\begin{document}

\title[{\sl XMM-Newton} observations of the Seyfert 1 AGN H\,0557$-$385]
    {{\sl XMM-Newton} observations of the Seyfert 1 AGN H\,0557$-$385}
\author[C.E.~Ashton et al.]
    {C.E.~Ashton$^1$\thanks{email: cea@mssl.ucl.ac.uk}, M.J.~Page$^1$, G.~Branduardi-Raymont$^1$,  A.J.~Blustin$^1$ \\
$^1$Mullard Space Science Laboratory, University College London, Holmbury St. Mary, Dorking, Surrey RH5 6NT\\}
\date{Accepted ??. Received ??}
\pubyear{2004}
\maketitle

\begin{abstract}

We present {\sl XMM-Newton} observations of the Seyfert 1 AGN H\,0557$-$385. We have conducted a study into the warm absorber present in this source, and using high-resolution RGS data we find that the absorption can be characterised by two phases: a phase with log ionisation parameter $\xi$ of 0.50 (where $\xi$ is in units of ergs cm s$^{-1}$) and a column of 0.2$\times10^{21}$ cm$^{-2}$, and a phase with log $\xi$ of 1.62 and a column of 1.3$\times10^{22}$ cm$^{-2}$. An iron K$\alpha$ line is detected. Neutral absorption is also present in the source, and we discuss possible origins for this. On the assumption that the ionised absorbers originate as an outflow from the inner edge of the torus, we use a new method for finding the volume filling factor. Both phases of H\,0557$-$385 have small volume filling factors ($\leq$ 1\%). We also derive the volume filling factors for a sample of 23 AGN using this assumption and for the absorbers with log $\xi$ $>$ 0.7 we find reasonable agreement with the filling factors obtained through the alternative method of equating the momentum flow of the absorbers to the momentum loss of the radiation field. By comparing the filling factors obtained by the two methods, we infer that some absorbers with log $\xi < 0.7$ occur at significantly larger distances from the nucleus than the inner edge of the torus.

\end{abstract}


\section{Introduction}

Warm absorbers are clouds of ionised gas intrinsic to AGN. They cause absorption at soft X-ray wavelengths, and nearby bright Seyferts are ideal for spectroscopic studies of this phenomenon. Warm absorbers are a useful probe of the inner environments of AGN, e.g. elemental abundances, outflow speeds, and degrees of ionisation. 

The importance of warm absorbers was underlined following a study by Reynolds (1997) which indicated that they are present in about half of all type 1 AGN. The advent of {\sl XMM-Newton} and {\sl Chandra} has shown the details of warm absorbers at their highest resolution yet, revealing that their primary spectroscopic signature is a series of narrow absorption lines (Kaastra et al 2000).

Many warm absorbers in type 1 AGN have now been observed with high resolution X-ray instruments, e.g. NGC~5548 (Kaastra et al 2000); IRAS~13349+2438 (Sako et al 2001); NGC~3783 (Kaspi et al 2002); IC~4329A (Steenbrugge et al 2005). Observations of warm absorbers typically show that there are at least two phases present, with distinctly different ionisation parameters, although in theory a continuum of ionisation parameters could be present between the observed phases. An unresolved transition array (UTA) of M-shell iron, first discovered in IRAS~13349+2438 (Sako et al 2001), has been found to be an important spectroscopic diagnostic tool (Behar et al 2001).

Two main theories exist as to where warm absorbers originate: either as evaporation from the obscuring molecular torus (e.g. Krolik \& Kriss 2001) invoked in Seyfert unification models (Antonucci 1993), or as an accretion disc wind (e.g. Elvis 2000). A detailed study of 23 AGN by Blustin et al (2005; hereafter B05) indicates that the warm absorbers of nearby Seyferts are most likely to originate as outflows from the molecular torus.

A recent hypothesis for the geometry of warm absorbers is that they have the form of an `ionisation cone': this scenario involves a cone of plasma, irradiated by the nuclear continuum, with a shape defined by the torus. From observations of the Seyfert 2 AGN NGC~1068, this scenario is suggested to account for both absorption features in Seyfert 1 AGN, and emission features in Seyfert 2 AGN (Kinkhabwala et al 2002; Brinkmann et al 2002; Ogle et al 2003).

H\,0557$-$385 (also known as 3A~0557-383, EXO\,055620-3820.2 and CTS B31.01) is a Seyfert 1 AGN at redshift 0.034 (Fairall, McHardy \& Pye 1982). The Galactic neutral column density towards it is 4$\times10^{20}$cm$^{-2}$ (Dickey \& Lockman 1990). We report the first observations of this source with {\sl XMM-Newton}. H\,0557$-$385 has also been observed with {\sl ASCA} (Turner et al 1996) and {\sl BeppoSAX} (Quadrelli et al 2003). Both these previous observations detected a warm absorber. In this paper we explore the details of the warm absorption in H\,0557$-$385, using both EPIC and RGS data.

\section{Observations and Data Reduction}
\label{Observations}

H\,0557$-$385 was observed with {\sl XMM-Newton} (Jansen et al 2001) on 3rd April 2002 and 17th September 2002; the observation details are summarised in Table~\ref{Obs}. 

\begin{table*}
\caption{Observation details. The RGS count rates are for the co-added first and second orders of RGS1 and RGS2, for each observation. The Effective Exposure Time is the exposure time with periods of high background excluded.}
\begin{tabular}{ccccccc}
\hline
Date Observed  & Effective Exposure Time & Background-subtracted Count Rate \\
&(ks)&(Counts s$^{-1}$) \\
\hline
03.04.2002 & pn: 8, MOS: 3, RGS: 12 & pn: 10.29$\pm{0.05}$, MOS1: 3.30$\pm{0.04}$, MOS2: 3.32$\pm{0.04}$, RGS:0.125$\pm{3\times10^{-3}}$    \\
17.09.2002 & pn: 7, MOS: 9, RGS: 8  & pn: 10.55$\pm{0.04}$, MOS1: 3.63$\pm{0.02}$, MOS2: 3.68$\pm{0.02}$, RGS:0.143$\pm{3\times10^{-3}}$     \\
\hline
\end{tabular}\\
\label{Obs}
\end{table*}

The EPIC pn data were taken in large window mode, and the MOS data in small window mode, both using the thin filter. We excluded periods of high background, which we identified in 5-10 keV lightcurves for the whole field of view outside the source region. The data were processed using the Science Analysis System (SAS) Version 6; the pn spectra were accumulated using single and double events, corresponding to PATTERN values of 0-4, and the MOS spectra were constructed using all valid events (PATTERN = 0-12). The count rates are below the thresholds where pile-up has to be considered (12 counts s$^{-1}$ for the pn large window mode, and 5 counts s$^{-1}$ for the MOS small window mode). Events next to bad pixels and next to the edges of CCDs were excluded (FLAG = 0 in SAS). 

The EPIC source spectra were constructed by selecting counts within a circle of radius 25 arcsec, and the background spectra were extracted from nearby source-free regions in circles of radius 75 arcsec. 

Response matrix files (RMF) and auxiliary response files (ARF) for each instrument were generated using the {\sl SAS}. The count rates for each observation are similar, as shown in Table~\ref{Obs}. The EPIC spectra from both observations were coadded using the method of Page, Davis \& Salvi (2003) and binned into 45 eV channels. 

The RGS spectra were extracted using {\sl rgsproc} in SAS V.6. The first and second order spectra and response matrices from both RGSs and both observations were resampled to the channels of the first order RGS1 spectrum from April 2002. The spectra were then coadded, and the response matrices were combined, to produce a single spectrum and a single response matrix. The spectrum was then binned by a factor of 9 to improve signal to noise, giving a binsize of 0.1 \AA\ at 20 \AA.

The data were analysed using {\sl SPEX} 2.00 (Kaastra et al 2002). In this paper we adopt the values H$_{0}$ = 70.0 km s$^{-1}$ Mpc$^{-1}$, $\Omega_{m}$ = 0.3 and $\Omega_{\Lambda}$ = 0.7.

\section{Results}
\label{Results}

\subsection{Fits to the EPIC data}

\begin{figure*}
\psfig{file=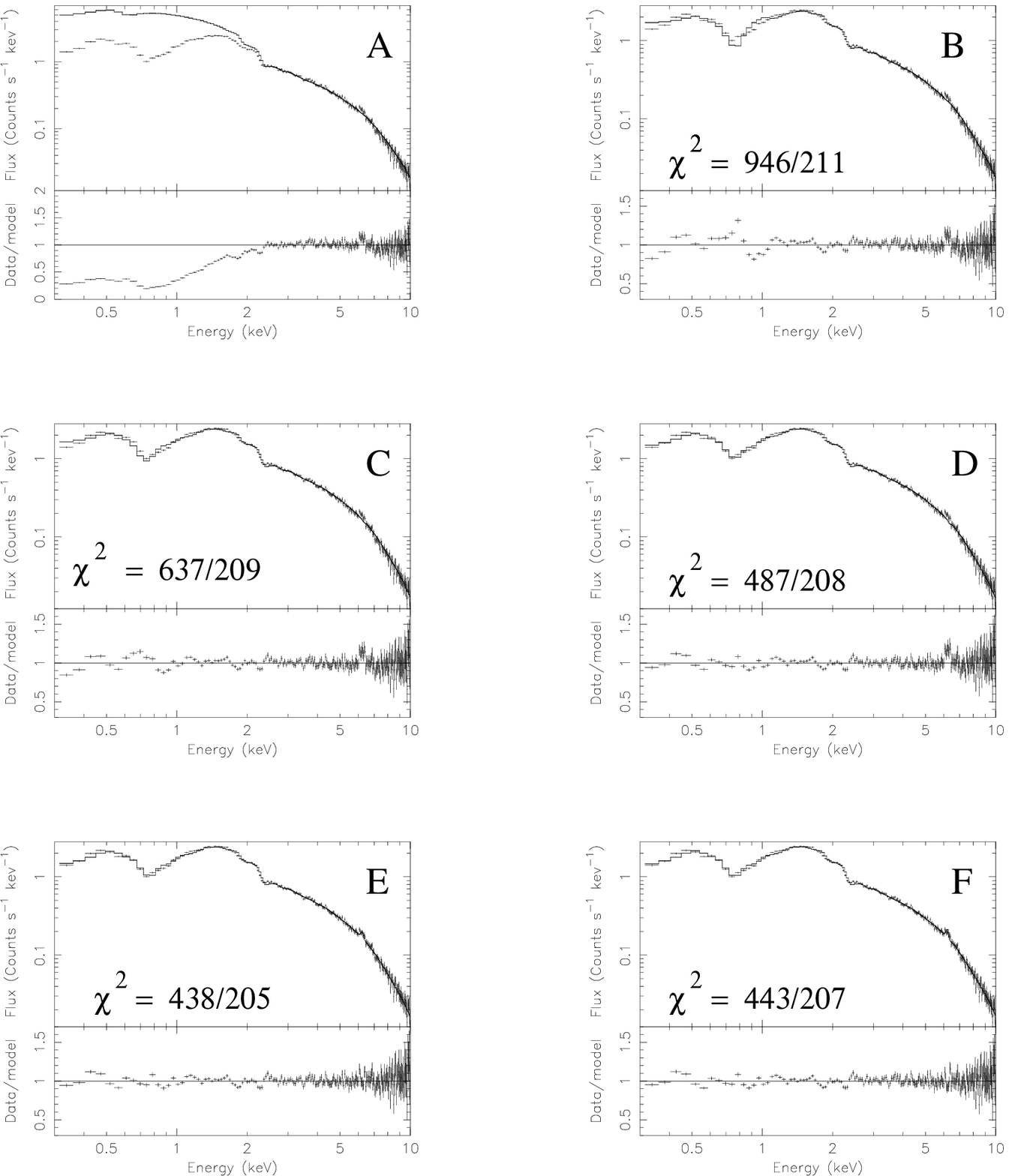,width=15cm,angle=360}
\caption{Spectral fits of EPIC data, plotted in the observed frame on logarithmic scales. The plots below each fit show the data/model ratio. A: Power law, fitted over 2.3-10 keV, plotted back to 0.3 keV. B: Power law and one warm absorber. C: Power law and two warm absorbers. D: Power law, two warm absorbers and neutral absorption. E: as D but including an iron K$\alpha$ line. F: as D but with a reflection component.}
\label{PN fits}
\end{figure*}

After initial inspection, the EPIC data were fitted in {\sl SPEX} with a number of spectral models: the fit parameters are listed in Table~\ref{Fits} and data and fitted models are shown in Fig.~\ref{PN fits}. Absorption by the Galactic column density is included in all the fits, using the model {\sl hot}, in {\sl SPEX}. This is for a gas in collisional ionisation equilibrium, set at the minimum temperature (5$\times10^{-4}$ keV), and it produces a very similar spectrum to that of a pure photoelectric model (e.g. Morrison \& McCammon 1983), but includes absorption lines as well as absorption edges.

We started by fitting a simple power law in the range 2.3-10 keV (Model A) to avoid the regions where the low-energy absorption affects the spectrum. In Fig.~\ref{PN fits}(A) the power law is plotted back to 0.3 keV to show the shape and extent of the absorption. The plot is suggestive of the presence of a warm absorber, with the deepest trough occuring at $\sim$ 0.7 keV; at this energy this will be a warm, rather than a neutral absorber.

To fit the whole spectrum, a warm absorber component ({\sl xabs} in {\sl SPEX}) was added to the model (Model B). The {\sl xabs} component models absorption by a cloud of photoionised gas, at a given column and ionisation parameter, where the ionisation parameter is defined as $\xi = L/nr^{2}$ in erg cm s$^{-1}$. Here $L$ is the 1-1000 Rydberg ionising luminosity (erg s$^{-1}$), $n$ the gas density (cm$^{-3}$), and $r$ the distance of the ionising source from the absorbing gas, in cm. We set the turbulent velocity of the gas to 100 km s$^{-1}$, and the elemental abundances to solar values. Only the column and ionisation parameter were left free in the fitting.

Including the warm absorber in the model brings it to resemble the data far more closely, as shown in the plot for Model B. However, the $\chi^{2}$/dof is 946/211, and the model is still not a good match to the data at some energies. The single {\sl xabs} component has created an absorption feature in the model which is deeper than that in the data at $\sim$ 0.75 keV, but the model has insufficent absorption at 1 keV, as examination of the data/model ratio in Model B shows.

Another {\sl xabs} component with a much higher ionisation parameter was added (Model C): this reduces the ionisation parameter and column of the first {\sl xabs} component and provides a better fit to the data at $\sim$ 0.75 keV and 1 keV, as shown in Fig.~\ref{PN fits}(C). The goodness of fit is significantly improved, but the model still appears to require some absorption at $\sim$ 0.3-0.4 keV to match the data. Therefore a component of neutral gas (again, using {\sl hot} in {\sl SPEX} at the minimum temperature of 5$\times10^{-4}$ keV) was added at the same redshift as H\,0557$-$385. The additional column density of neutral gas, 4$\times10^{20}$ cm$^{-2}$, provides a much better fit to the data at lower energies, as shown in Fig.~\ref{PN fits}(D). With the addition of neutral gas, the column of the high-$\xi$ absorber increases and the model fits the data more closely at $\sim$ 0.3-0.4 keV and $\sim$ 0.5-1 keV.

Now the soft X-ray spectrum is fitted well, we turn to the hard X-ray spectrum. An iron K$\alpha$ line is visible at $\sim$ 6 keV and we first modelled it with a gaussian. This improved the $\chi^{2}$ by 49 for 3 extra parameters (Model E in Fig.~\ref{PN fits}). The energy of the line is 6.41$\pm{0.05}$ keV, with a FWHM of 330$^{+170}_{-130}$ eV and equivalent width of 85$^{+33}_{-24}$ eV. As this line may originate by fluorescence in reflecting material, we next tried a reflection model ({\sl refl} in {\sl SPEX}), which includes an iron K$\alpha$ line, in combination with Model D. As the energy of the line indicates it comes from cold material, we assume the reflecting material is cold. The best-fit reflected fraction is 0.39$^{+0.11}_{-0.10}$. The addition of this component improves the $\chi^{2}$ by 44 from Model D. This is shown in Fig.~\ref{PN fits} as Model F. As the reflected fraction is small, and the addition of this component produces a slightly worse $\chi^{2}$ than Model E, we conclude that the inclusion of reflection is not justified and that the iron K$\alpha$ line alone is a sufficient addition to Model D. We then take Model E as the best-fit to the EPIC data.

The 2-10 keV flux of model E is 4.3$\pm{0.1}\times10^{-11}$ erg s$^{-1}$ cm$^{-2}$, compared to 2$\times10^{-11}$ erg s$^{-1}$ cm$^{-2}$ at the time of the {\sl ASCA} observation of Turner et al (1996) and 4$\times10^{-11}$ erg s$^{-1}$ cm$^{-2}$ at the time of the {\sl BeppoSax} observation of Q03. The 1-1000 Rydberg luminosity of model E is 4.3$\times10^{44}$ erg s$^{-1}$.

\begin{table*}
\caption{: Parameters of fits to the combined EPIC data. The Galactic column (4$\times10^{20}$ cm$^{-2}$) is included in the models. PL norm is the normalisation of the power law in units of $10^{52}$ ph s$^{-1}$ keV$^{-1}$. $\xi$ is the ionisation parameter in units of erg cm s$^{-1}$. The reflection fraction is defined as $\Omega$/2$\pi$, where $\Omega$ is the solid angle subtended by the reflector at the X-ray source. Errors are at 90 \% confidence for one interesting parameter ($\Delta \chi^{2} = 2.71$).}

\vspace{0.4cm}

\scriptsize
\begin{tabular}{c@{\hspace{2mm}}c@{\hspace{2mm}}c@{\hspace{2mm}}c@{\hspace{2mm}}c@{\hspace{2mm}}c@{\hspace{2mm}}c@{\hspace{2mm}}c@{\hspace{2mm}}c@{\hspace{2mm}}c@{\hspace{2mm}}c@{\hspace{2mm}}c}
Model & $\Gamma$ & PL norm & Warm absorber  & log $\xi$ of & Neutral N$_{\mbox{\sc h}}$ & Fe K$\alpha$ & Fe K$\alpha$ & Fe K$\alpha$ & Reflection  & $\chi^{2}$/d.o.f   \\   

&& at 1 keV  & N$_{\mbox{\sc h}}$  ($10^{22}$cm$^{-2}$) & warm absorber & ($10^{22}$cm$^{-2}$) & energy (keV) & $\sigma$ (eV) & EW (eV) & fraction    \\ 
\hline

\\

PL$\times$xabs & 1.79$\pm{0.01}$ & 6.91$\pm{0.13}$ & 0.76$\pm{0.02}$ & 0.41$\pm{0.02}$  & --  & -- & -- & -- & -- & 946/211   \\          
(Model B)

\\
\\

PL$\times$xabs$\times$xabs & 1.83$\pm{0.01}$ & 7.48$\pm{0.20}$ & 0.62$\pm{0.02}$ & 0.07$\pm{0.05}$  & -- & -- & -- & -- &  -- &  637/209  \\
(Model C) &&&   0.75$\pm{0.10}$ & 2.25$\pm{0.06}$

\\
\\

PL$\times$xabs$\times$xabs & 1.88$\pm{0.01}$ & 8.11$\pm{0.20}$ & 0.60$\pm{0.02}$  & 0.37$\pm{0.05}$  & 0.04$\pm{0.01}$ & -- & -- & -- & -- &  487/208  \\

$\times$neutral N$_{\mbox{\sc h}}$(Model D) &&&  1.11$\pm{0.20}$  & 2.30$\pm{0.07}$

\\
\\

(PL+Fe K$\alpha$)$\times$xabs$\times$xabs  & 1.90$\pm{0.02}$ & 8.37$\pm{0.20}$ & 0.61$\pm{0.02}$ & 0.36$\pm{0.04}$   & 0.05$\pm{0.01}$  & 6.41$\pm{0.05}$ &  140$^{+70}_{-50}$  & 85$^{+30}_{-20}$  & -- & 438/205   \\

$\times$neutral N$_{\mbox{\sc h}}$(Model E) &&&  1.27$\pm{0.20}$ & 2.33$\pm{0.05}$

\\
\\

(PL+refl)$\times$xabs$\times$xabs & 1.93$\pm{0.02}$ & 8.60$\pm{0.30}$ & 0.61$\pm{0.02}$  & 0.35$\pm{0.04}$  & 0.05$\pm{0.01}$ & -- & -- & -- & 0.39$\pm{0.10}$ &  443/207  \\

$\times$neutral N$_{\mbox{\sc h}}$ (Model F) &&&   1.24$\pm{0.20}$  & 2.31$\pm{0.06}$ 

\\
\\

\hline
\end{tabular}
\label{Fits}
\end{table*}

\subsection{Fits to the RGS data}

We now investigate the soft X-ray absorption features using the RGS, which has much higher energy resolution (E/$\Delta$E $\sim$ 200-800) than EPIC (E/$\Delta$E $\sim$ 20-50).

\begin{figure*}
\psfig{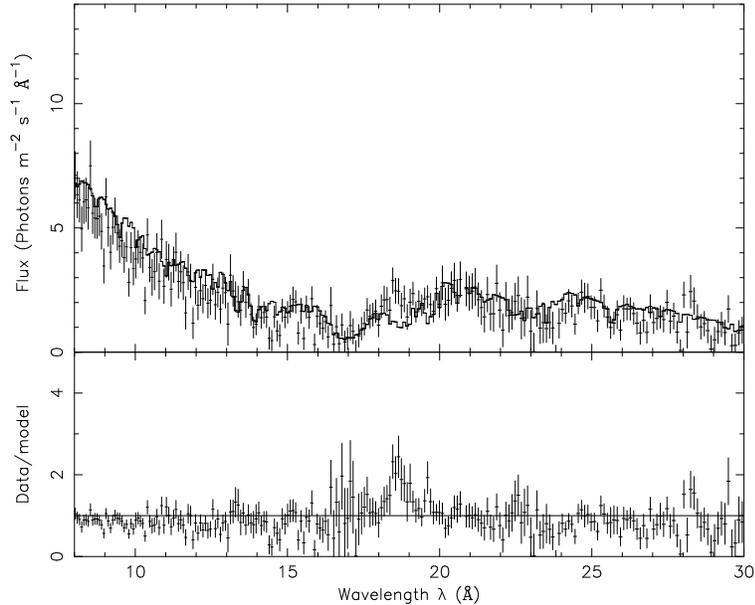}
\caption{The RGS spectrum, plotted in the observed frame with the best-fit EPIC model E superimposed. The data/model ratio is also shown.}
\label{rgs_epic_xabs}
\end{figure*}

\begin{figure*}
\psfig{file=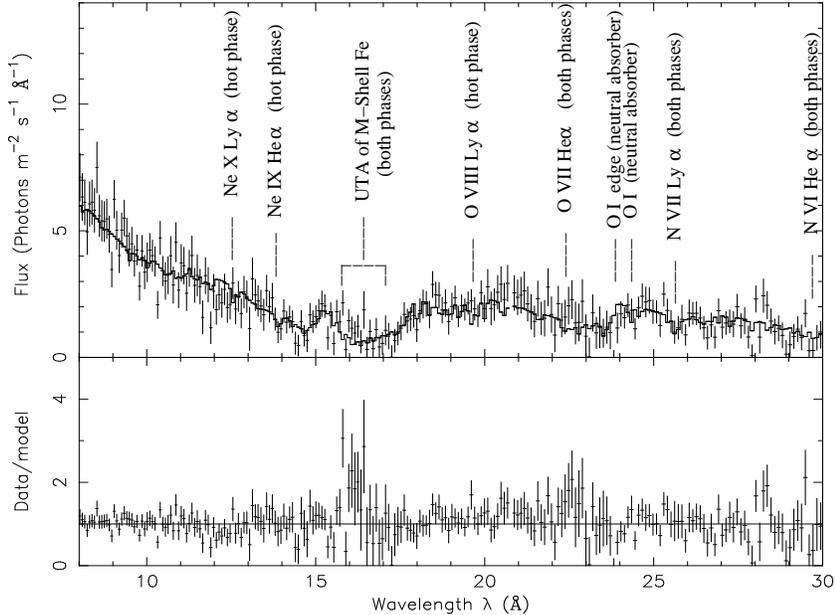,width=11cm,angle=0}
\caption{The RGS spectrum, plotted in the observed frame with the best-fit RGS model. Some of the main absorbing ions that belong to each phase are labelled. The data/model ratio is also shown.}
\label{rgs_xabs_fig}
\end{figure*}

We plotted the EPIC best-fit, model E, on to the RGS spectra after convolving it with the RGS response; the result is shown in Fig.~\ref{rgs_epic_xabs}. RGS and EPIC cross-calibration shows discrepancies of up to 15 per cent in the band 6-18 \AA\, and up to 10 per cent in $\sim$ 18-36 \AA\, (Fig 11 of Kirsch et al 2004). The spectral shape and flux of H\,0557$-$385 are generally well reproduced, but this model does not match the RGS data completely. It overestimates the data over $\sim$ 8-15 \AA\, and $\sim$ 23-33 \AA\, and is too absorbed at $\sim$ 18-19 \AA\,. In order to obtain a better fit to the RGS data, we kept the continuum from EPIC model E fixed, and the columns and ionisation parameters of the absorbers in the model were left to vary. We consider only the 8-30 \AA\, spectrum, where RGS calibration and statistical quality of the data are best. The turbulent velocity of the gas was again set to 100 km s$^{-1}$. The absorbers were set at the recession velocity of the AGN. The resultant best-fit proved to be in much better agreement with the RGS data, and is shown in Fig.~\ref{rgs_xabs_fig}, along with the data/model ratio. The main ions from each absorbing phase are indicated. The parameters from the EPIC model E and RGS fits are compared in Table.~\ref{RGS xabs}.

As the RGS has better energy resolution than EPIC, the best-fit model will be closer to the real characteristics of the source. As shown in Table.~\ref{RGS xabs}, the parameters of the warm absorbers required by the RGS data are different to those from the EPIC data. With respect to the EPIC data, for the RGS data the column of the low-$\xi$ phase is lower whereas the column of the high-$\xi$ phase has not changed. The log $\xi$ values of the low-$\xi$ phase are consistent, but the log $\xi$ of the high-$\xi$ phase is smaller than that derived from the EPIC data. A much larger column of neutral gas is also required. The discrepancy between the EPIC and RGS column densities for the neutral and low-$\xi$ absorbers could be related to the larger uncertainty in the pn response at the lowest energies.

Compared to Fig.~\ref{rgs_epic_xabs}, the model in Fig.~\ref{rgs_xabs_fig} is now a much better fit to the data at 8-15 \AA\, and from 17 \AA\, onwards. The $\chi^{2}$/dof of the model shown in Fig.~\ref{rgs_epic_xabs} is 413/236, whereas the $\chi^{2}$/dof of the RGS best-fit model (Fig.~\ref{rgs_xabs_fig}) is 246/231. The discrepancy between the data and the model at $\sim$ 18-19 \AA\, in Fig.~\ref{rgs_epic_xabs} has disappeared.

\begin{figure*}
\psfig{file=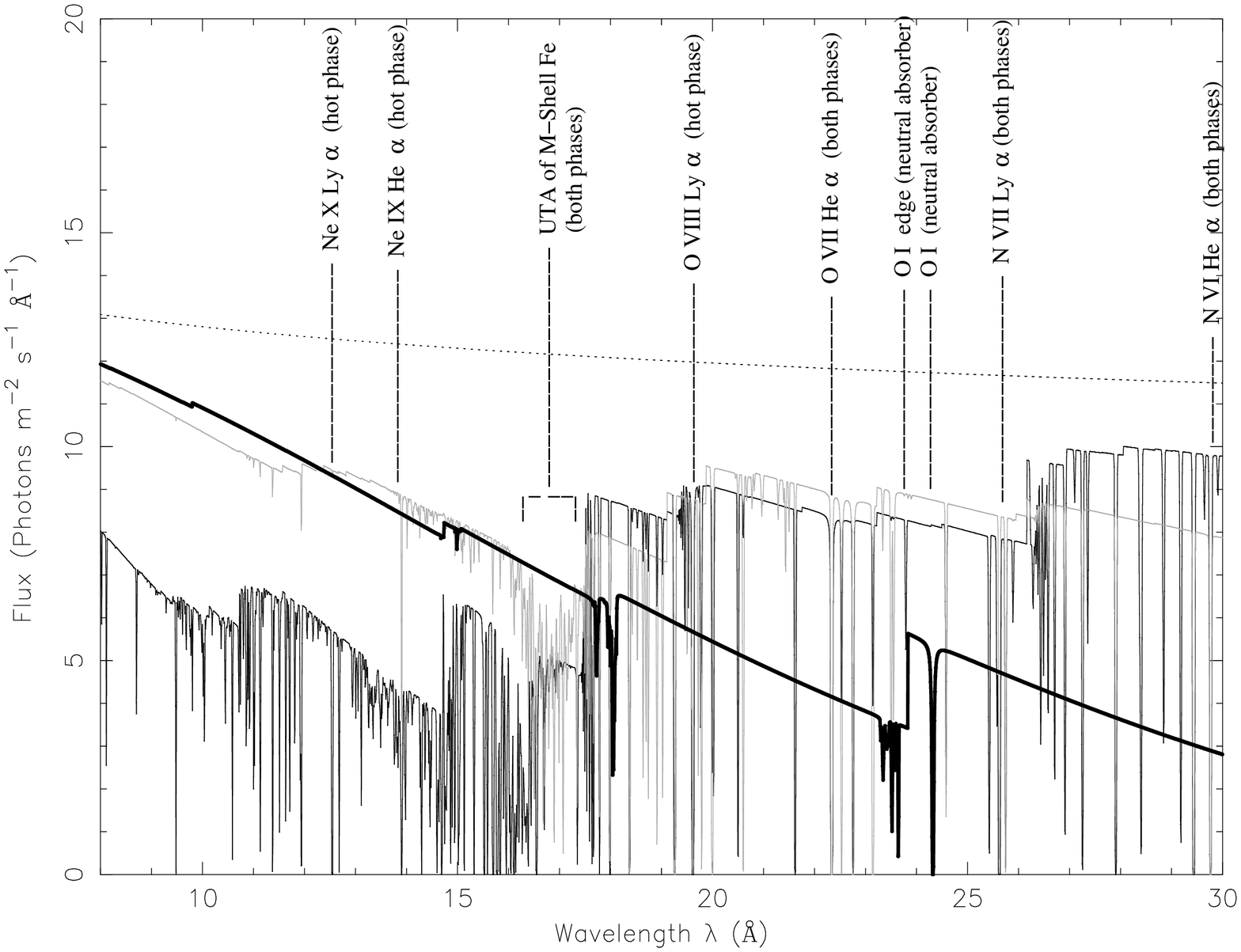,width=10cm,angle=0}
\caption{The RGS best-fit model, plotted in the observed frame, for the two absorber phases. Black: high-$\xi$ phase. Grey: low-$\xi$ phase. The top dotted curve is the unabsorbed continuum; the lower curve, shown in bold black, includes the neutral gas. The Galactic column is not included in any of the models in this plot.}
\label{Colour_models}
\end{figure*}

Fig.~\ref{Colour_models} shows the models corresponding to the high-$\xi$ and low-$\xi$ phases. The deep and broad absorption feature at $\sim$ 16-18 \AA, clearly visible in the RGS spectrum, is the signature of an unresolved transition array (UTA) of M-shell iron. The log $\xi$ values of the two phases are such that a UTA is present in both of them. There is much more absorption at shorter wavelengths for the high-$\xi$ phase than for the low-$\xi$ phase, as expected.

To check the self-consistency of the RGS-derived absorption parameters
with the continuum from the EPIC fit, we re-fitted the EPIC spectrum
with the absorption parameters fixed at the RGS values.  The best fit
continuum changes from $\Gamma$=1.90 to $\Gamma$=1.96, consistent at 3
sigma; the difference is well within the absolute uncertainty in the
EPIC calibration, and is much less than the typical difference
resulting from the EPIC-RGS cross calibration.

\begin{table*}
\caption{Parameters of fits to the RGS data, compared with EPIC model E. Errors are at 90 \% confidence ($\Delta \chi^{2} = 2.71$) for one interesting parameter.}
\vspace{0.4cm}
\begin{tabular}{cccccccccccc}
\hline
Instrument  & Warm N$_{\mbox{\sc h}}$  & log $\xi$ & Neutral N$_{\mbox{\sc h}}$ & $\chi^{2}$/dof  \\
& ($10^{22}$cm$^{-2}$) && ($10^{22}$cm$^{-2}$)   \\ 
\hline 

EPIC & 0.61$\pm{0.02}$, 1.27$\pm{0.20}$ & 0.36$\pm{0.04}$, 2.33$\pm{0.05}$  & 0.05$\pm{0.01}$ & 438/205   \\

RGS & 0.20$\pm{0.10}$, 1.31$\pm{0.20}$  & 0.50$\pm{0.30}$, 1.62$\pm{0.10}$ & 0.12$\pm{0.02}$ & 246/231   \\

\hline
\end{tabular}\\
\label{RGS xabs}
\end{table*}

\section{Discussion}
\label{Discussion}

The EPIC data tell us broadly the characteristics of the warm absorbers; model E tells us that the warm absorption is caused by two phases with different ionisation parameters. A column of neutral gas is also required to explain the spectrum at $\sim$ 0.3-0.4 keV. The RGS data, at higher resolution, require a smaller log $\xi$ value for the high-$\xi$ phase and a smaller column for the low-$\xi$ phase. A larger column of neutral gas is also required for the RGS spectrum. 

The Fe K$\alpha$ line, with an energy of 6.41$\pm{0.05}$ keV, originates in material that could be neutral, or ionised up to ionisation of Fe XVIII, which is consistent with this energy (Kallman et al 2004). This wide range of ionisation states corresponds to log $\xi$ $<$ 2.3, so we cannot constrain well the ionisation state of the line-emitting material. Assuming it is a single line, it is resolved in EPIC with a FWHM of 16,000$^{+8000}_{-6000}$ km s$^{-1}$, whereas the {\sl ASCA} observation of Turner et al (1996) yielded a FWHM of 55,000$^{+22,000}_{-24,000}$ km s$^{-1}$. However, the width of the line could be due to contributions from multiple ionic species of iron. We find the equivalent width of the line to be 85$^{+33}_{-24}$ eV, compared with $\sim$ 300 eV found by Turner et al (1996).

From optical spectroscopy, Rodr\'{i}guez-Ardila et al (2000) distinguish 3 H$\alpha$ components in H\,0557$-$385. The FWHM of the narrowest component is 1035 km s$^{-1}$, that of the intermediate component 2772 km s$^{-1}$ and for the broadest component it is 11,000 km s$^{-1}$. The last value suggests that the region that produces the broadest H$\alpha$ component could also be the location of the Fe K$\alpha$ line emitting gas, if this latter emission is due to a single line.

To start to build up a picture of the AGN as a whole, we use the relation in Krolik \& Kriss (2001) to estimate the approximate distance to the inner edge of the torus due to dust sublimation and photoionisation, $R_{torus}$, in cm, from the continuum source with ionising luminosity $L$, in erg s$^{-1}$:

\begin{equation}
R_{\rm torus, \hspace{1mm}cm}  \sim  3\times10^{-4} \sqrt{L}   
\label{equation2}
\end{equation}

Using the 1-1000 Rydberg luminosity from the EPIC best-fit model, i.e. $L$ = 4.3$\times10^{44}$ erg s$^{-1}$, we obtain 6.2$\times10^{18}$ cm ($\sim$ 2 pc) for the inner radius of the torus.

To place a distance on the Fe K$\alpha$ line formation region from the black hole, we use a black hole mass of 3$\times10^{7}$ M$_{\odot}$. This value is intermediate to the Schwarzchild and Kerr black hole masses measured for H\,0557$-$385 by Rokaki \& Boisson (1999), who assume that H$\beta$ emission lines come from the disc. We use the FWHM of 16,000 km s$^{-1}$ of the Fe K$\alpha$ line as $v$ in the equation:
\begin{equation}
v  = \sqrt\frac{GM}{R}
\label{equation3}
\end{equation}
to obtain the distance, $R$, of the line formation region from the black hole, assuming the line emitting gas moves in a Keplarian orbit: we obtain a distance of 2$\times10^{15}$ cm. The Schwarzchild radius for the black hole, 2GM/c$^{2}$, will be 9$\times10^{12}$ cm. If we take the inner edge of the accretion disc to be 6GM/c$^{2}$, then for H\,0557$-$385 this is 3$\times10^{13}$ cm. For the H$\alpha$ line we find distances of 4$\times10^{17}$ cm, 5$\times10^{16}$ cm, and 3$\times10^{15}$ cm for the narrowest, intermediate and broadest components respectively.

\subsection{Where did the warm absorbers originate?}

We now try to determine the distances of the warm absorbers from the continuum source, and thus find some clue as to their origins.  

We assume the warm absorbers to be a continous, constant velocity wind. We use R$_{l}$ as the launching radius of the wind; for such a wind, N$_{\mbox{\sc h}} = n_{l}R_{l} f$ and $\xi = L/n_{l}R_{l}^{2}$. $L$ is the 1-1000 Rydberg luminosity (4.3$\times10^{44}$ erg s$^{-1}$ from the EPIC best-fit model), $n_{l}$ is the density at $R_{l}$ and $f$ is the volume filling factor of the gas. Substituting to eliminate $n_{l}$, we obtain

\begin{equation}
L = \frac {\xi N_{\mbox{\sc h}} R_{l}} {f}
\label{equation4}
\end{equation}

Taking the upper limit of the volume filling factor, $f = 1$, and including the 90 per cent confidence limits on the other parameters, we obtain upper limits of $R_{l}$ $\leq$ 1$\times10^{21}$ cm to the higher-$\xi$ absorber, and $R_{l}$ $\leq$ 3$\times10^{23}$ cm to the lower-$\xi$ absorber. Unfortunately, the locations of the warm absorbers cannot be constrained any better from the data available.

\subsection{A torus origin for warm absorbers?}
\label{A torus origin for warm absorbers?}

From the relation for the ionisation parameter, $\xi = L/nr^{2}$, we infer that for constant density and $\xi$, $r$ scales with L$^{1/2}$. Equation~\ref{equation2} expresses the same scaling relation, for the inner edge of the torus. Here, $r$ is the distance of the ionising source from the absorbing gas. From this we can infer that if warm absorbers originate at the inner edge of the torus then they should possess similar values of the product of ionisation parameter and density, independent of their ionising luminosities. 

We re-write Equation~\ref{equation2} as $R_{\rm torus}$ $\sim$ $K\sqrt{L}$ with $K$ = 3$\times10^{-4}$. Then re-arranging Equation~\ref{equation4}, equating $R_{l}$ to $R_{\rm torus}$, and substituting for $R_{\rm torus}$, we obtain

\begin{equation}
L^{1/2} \simeq \frac {\xi N_{\mbox{\sc h}} K} {f}
\label{equation5}
\end{equation} 

Taking logs of both sides then working through the algebra, we obtain

\begin{equation}
log \, L \simeq 2 \, \log \, {\xi N_{\mbox{\sc h}}} - 2 \, \log \frac {f}{K}
\label{equation6}
\end{equation}

\begin{figure*}
\psfig{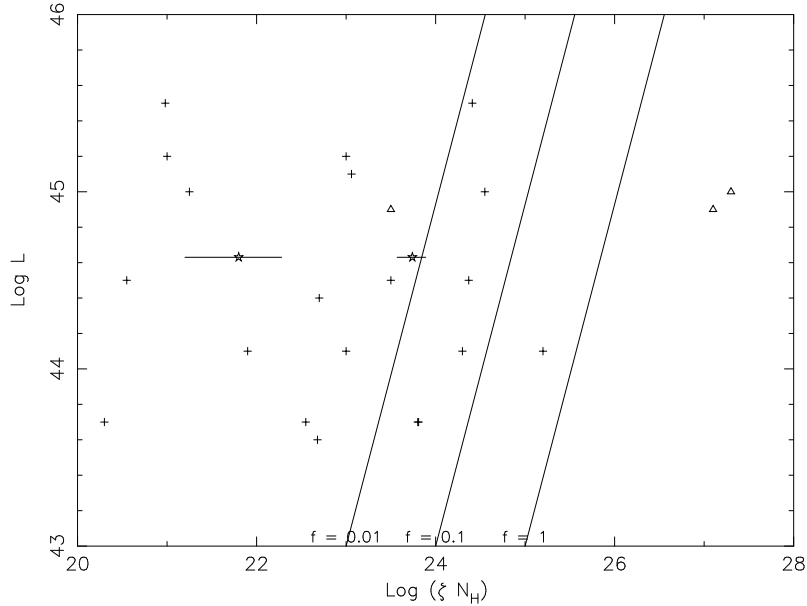}
\caption{Log values of ionising luminosity, versus the log of the product of ionisation parameter and column density, for 14 AGN. H\,0557$-$385 is shown as starred points, along with the errors. Data for all other AGN are from B05. The warm absorbers in PG~0844+349 and PG~1211+143, believed to be accretion disk winds, are shown as triangles. Some different filling factors are labelled.}
\label{Torus_orig}
\end{figure*}

This implies a linear relationship between $\log \, L$ and $\log \, {\xi N_{\mbox{\sc h}}}$ with a gradient of 2 and an intercept of $-2 \, \log \frac {f}{K}$. A list of $\log N_{\mbox{\sc h}}$, $\log \xi$ and $\log L$ values for warm absorbers observed in 14 AGN is given in Tables 2 and 4 of B05. These authors assume a constant velocity wind in considering the energetics of warm absorbers. In Fig.~\ref{Torus_orig} we plot log $L$ versus $\log \, {\xi N_{\mbox{\sc h}}}$  for the AGN in B05 along with the parameters we derive for H\,0557$-$385 from the RGS data.

We can use Equation~\ref{equation6} to calculate the filling factor $f$ when all the other terms are known. The values of $L$, $N_{\mbox{\sc h}}$ and $\xi$ have been determined from our observations of H\,0557$-$385 and are listed in B05 for 14 other AGN. The filling factors for each absorber phase are listed in Table~\ref{filling_factors}.

The absorber of PG~0844+349 and one of the absorbers in PG~1211+143 would have filling factors larger than 1 if they originate at the inner edge of the torus, so they cannot originate as torus winds. This would also hold if we consider accelerating rather than constant velocity winds, as in this case N$_{\mbox{\sc h}} < n_{l}R_{l} f$ and the apparent value of $f$ would be larger than that indicated in Table~\ref{filling_factors}. Pounds et al (2003a, 2003b) and B05 argue that the absorbers in these objects originate as accretion disk winds; from Fig.~\ref{Torus_orig} these must originate closer than the torus, so the results are consistent with an accretion disk wind.

From Fig.~\ref{Torus_orig} and Table~\ref{filling_factors} there are 6 absorbers for which we compute very low filling factors ($f < 10^{-4}$). These absorbers have the smallest values of log $\xi$ in the sample, ranging from 0 to 0.68. The AGN in question are H\,0557$-$385, MR~2251-178, IRAS~13349+2438, NGC~4593, NGC~5548 and Ark~564.

\begin{figure*}
\psfig{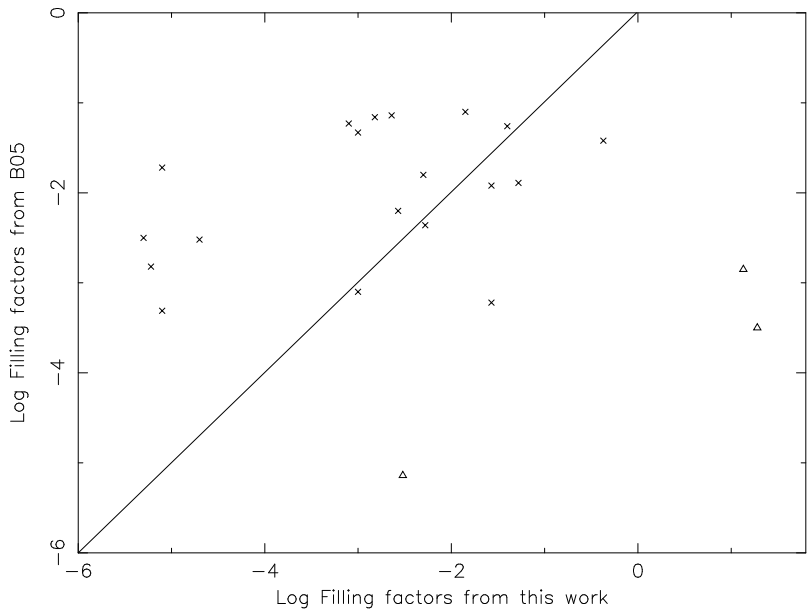}
\caption{The filling factors from this paper plotted versus the filling factors derived from B05, which are translated from percentages into a fractional scale. The warm absorbers in PG~0844+349 and PG~1211+143, believed to be accretion disk winds, are shown as triangles. A line indicating equivalence between our results and those of B05 is shown.}
\label{alex_ceri}
\end{figure*}

In Fig.~\ref{alex_ceri} we plot the filling factors for the AGN as listed in Table~\ref{filling_factors}, versus those found by BO5. The advantage of our method for calculating filling factors compared to that of B05 is that our method only requires the $\log N_{\mbox{\sc h}}$, $\log \xi$ and $\log L$ values for warm absorbers, and no dynamical information is required. In contrast, the method of B05 used the outflow velocities of the warm absorbers, and involves calculating the momentum absorbed and scattered by the warm absorber. So two different methods exist for finding the filling factor of the warm absorber.

Unsurprisingly, Equation~\ref{equation6} assigns much larger filling factors to the accretion disk wind absorbers than B05. For the low-$\xi$ absorbers, the two methods produce inconsistent results. For the remaining objects, the absorbers with log $\xi > 0.7$ are located in a different part of Fig.~\ref{alex_ceri} to the 5 lower-$\xi$ absorbers, which are found on the left-hand side of the plot. The absorbers with log $\xi > 0.7$ have -4 $< \log f <$ 0 in both methods. Their mean log $f$ from B05 is -1.8$\pm{0.2}$ with a standard deviation of 0.7, and their mean log $f$ from this work is -2.1$\pm{0.2}$ with a standard deviation of 0.8 where the errors are 1 $\sigma$. Therefore the two methods yield consistent values for $\langle \log f \rangle$. It is worth comparing the standard devations of these results with the uncertainties inherent in the two methods. For our method we estimate the overall uncertainty on $f$ to be approximately 0.5 dex (a sum in quadrature of 0.15 dex from L$_{ion}$, 0.2 dex from $\xi$, 0.3 dex from N$_{\mbox{\sc h}}$ and 0.3 dex from $K$, the latter from the uncertainty in the distance of the inner edge of the torus). For the method of B05, we estimate an uncertainty on $f$ of approximately 0.6 dex (a sum in quadrature of 0.3 dex from L$_{ion}$, 0.2 dex from $\xi$ and 0.5 dex from the absorbed and scattered momentum). Hence, the standard deviations on $f$ derived from the two methods are comparable with the errors we estimate for the two methods. For the absorbers with log $\xi < 0.7$, the mean log $f$ from our method is -5.1$\pm{0.1}$ and that of B05 is -2.6$\pm{0.3}$.

If we relax our assumption that the absorbers originate at the inner edge of the torus, we can determine the values of $R_{l}$ that would bring the two methods into agreement, by substituting the filling factors of B05 into Equation~\ref{equation4}. These values of $R_{l}$ are given in column 8 of Table~\ref{filling_factors}. Taking the standard deviations on $f$ found from the high-$\xi$ group of absorbers as the uncertainties inherent in the two methods, and summing them in quadrature, we obtain an uncertainty of $\sim$ 1 dex. This explains why some of the values of $R_{l}$ determined are very large (the low-$\xi$ absorbers in MR~2251-178 and Ark~564 are estimated to be greater than 3 Kpc distant from the nucleus). The mean of log $R_{l}$ for the low-$\xi$ absorbers is 2.63, so a typical distance for these absorbers is $\sim$ 400 pc, which is similar to the scale of the narrow-line region (Osterbrock 1993, Capetti et al 1999). At a similar distance from the continuum source, a photoionised cloud is seen in emission in X-ray observations of the prototypical Seyfert 2 galaxy NGC~1068 (Brinkman et al 2002, Ogle et al 2003). The low-$\xi$ absorbers could be equivalent to this photoionised cloud in NGC~1068, but seen in absorption rather than in emission. It is unclear whether such absorbers originate from the torus or from other, more distant, material.

\begin{table*}
\caption{Comparison of volume filling factors (f) between this work, and B05. The B05 filling factors are translated from percentages into a fractional scale. `w.a.' stands for warm absorber. $R_{torus}$ is the distance of the inner torus edge, and $R_{l}$ is the launch radius calculated for the warm absorbers using the filling factors from B05. The uncertainty on R$_{l}$ is approximately $\pm{1}$ dex, see Section.~\ref{A torus origin for warm absorbers?}.}
\vspace{0.4cm}
\begin{tabular}{ccccccccccc}
\hline
AGN  & Log $\xi$ & Log N$_{\mbox{\sc h}}$  & Log L$_{ion}$ & f,         & f,        &  R$_{torus}$  & R$_{l}$       \\
     & of w.a.     & of w.a.   &                     & this work  & from B05  &  pc           & pc             \\   
\hline

H\,0557$-$385 & 0.50 &  21.3  & 44.6  & 9$\times10^{-5}$ & ---                 & 2.0          &   ---     \\
            & 1.62 &  22.1  &         & 0.008              & ---               &              &   ---     \\

\\

MR~2251-178 & 2.9 &  21.51  & 45.5  & 0.014            & 0.079                 &  5.5         &  30                    \\ 
	    & 0.68 & 20.3  &       & 5$\times10^{-6}$  & 0.0031                &              &  3000                  \\	

\\

PG~0844+349 & 3.7 & 23.6  & 45.0  & 18.93              &  3.1$\times10^{-4}$   & 3.1          &  5$\times10^{-5}$     \\
 
\\

PG~1211+143 & 3.4 & 23.7  & 44.9  & 13.4               &  0.0014               & 2.7          &  3$\times10^{-4}$      \\
            & 1.7 & 21.8  &       & 0.003              &  7.2$\times10^{-6}$   &              &  0.006                 \\

\\

IRAS~13349+2438 & 0 & 21.25   & 45.0  & 2$\times10^{-5}$ & 0.0030              & 3.1          &  500          \\

\\

NGC~4593 & 2.61 & 21.2  & 43.7  & 0.027               & 0.012                  &  0.69        &  0.30           \\
	 & 0.5  & 19.8  &       & 8$\times10^{-6}$    & 4.9$\times10^{-4}$     &              &  40              \\	

\\

NGC~3783 & 1.1 & 21.9   & 44.1  & 0.0027              & 0.0063                &   1.1         &  3            \\
	 & 2.3 & 22.0   &       & 0.053               & 0.013                 &               &  0.3          \\              
 	 & 2.9 & 22.3   &       & 0.424               & 0.038  	              &               &  0.1           \\

\\

Mkn~509  & 1.76 & 21.3  & 45.1  & 0.001               & 0.047 	              &  3.4          &  200          \\

\\

NGC~7469  & 2.1  & 20.6  & 44.4  & 0.001              & 8.6$\times10^{-4}$    &  1.5          &  1          \\

\\

NGC~3516  & 0.78 & 21.9  & 43.6   & 0.0023            & 0.072                 &  0.6          &  20            \\

\\

NGC~5548  & 2.69  & 21.68 & 44.5   & 0.040            & 0.055                 &  1.7          &  2         \\
	  & 1.98  & 21.52 &        & 0.005            & 0.016                 &               &  5     	\\
	  & 0.4   & 20.15 &        & 6$\times10^{-6}$ & 0.0015                &               &  400          \\	

\\

MCG -6-30-15  & 1.25  & 21.3  & 43.7   & 0.0015       & 0.069                 &  0.69         &  30          \\
	      & 2.5   & 21.3  &        & 0.027  & 6$\times10^{-4}$            &               &  0.02         \\

\\

NGC~4051  & 1.4  & 21  & 42.3          & 0.0053       & 0.0044                & 0.14          &  0.1  	 \\

\\

Ark~564  & 0   & 21  & 45.2   & 8$\times10^{-6}$ & 0.019                      & 3.9           &  10000         \\
	 & 2   & 21  &        & 8$\times10^{-4}$ & 0.059                      &               &  300      	\\	

\hline

\end{tabular}\\
\label{filling_factors}
\end{table*}

\subsection{Where is the neutral gas component?}

We find that the RGS data of H\,0557$-$385 require absorption from neutral gas local to the galaxy, of column density 1.2$\times10^{21}$cm$^{-2}$. But where is this gas in the AGN structure? Fairall, McHardy \& Pye (1982) and Rafanelli (1985) found the optical emission lines of H\,0557$-$385 heavily reddened. Therefore if we assume that dust is mixed with the neutral absorber, then both must lie outside the broad line region. There are therefore three main possibilities for the location of the neutral gas: the AGN itself, the disk of the host galaxy, or somewhere in the host galaxy but outside the disk.

{\sl Scenario 1}: The first possibility is gas within the AGN itself. If the warm absorber is a wind formed by evaporation from the dusty torus, then the neutral gas could be from the torus. Ogle (2003) suggests that the torus has a role in defining the shape of the ionisation cone. As we must be viewing the continuum source through the ionisation cone to see a warm absorber, we cannot be looking directly through the side of the torus. However, it could be that we are seeing cold absorption through a small cross-section of the torus if our line of sight grazes the edge.

{\sl Scenario 2}: The second possibility is the disk of the host galaxy. H\,0557$-$385 is classified as S0a in the Nasa Extragalactic Database, between a spiral and a lenticular galaxy. It is observed at an inclination angle $i$ = 75$^\circ$ from the normal (Crenshaw \& Kraemer 2001). If the host galaxy of H\,0557$-$385 has similar column density values to other Sa galaxies, then the neutral absorption we observe could originate in the host galaxy of the AGN. The average neutral hydrogen column density for an Sa galaxy viewed face-on, $N_{\mbox{\sc h}\perp}$, is 1.4$\times10^{21}$ cm$^{-2}$ (Broeils \& Woerden 1994). Using the relation $N_{\mbox{\sc h}} = N_{\mbox{\sc h}\perp}/$cos~$i$, we obtain an expected value of 5.3$\times10^{21}$ cm$^{-2}$ for $N_{\mbox{\sc h}}$. This is larger than the neutral column we observe in H\,0557$-$385, so it is plausible the neutral gas in H\,0557$-$385 could originate in the disk of the host galaxy.

{\sl Scenario 3}: The third possibility for the origin of this neutral gas is outside the AGN and above the disk of the galaxy, but within the structure of the galaxy. It could be in the form of a dust lane. From HST images, Malkan, Gorjian \& Tam (1998) find that there are dust lanes at distances of hundreds of parsecs in Seyfert 1 and 2 AGN. These dust lanes have little or no connection with the central engine and could have column densities up to $\sim$ 10$^{23}$ cm$^{-2}$ (Matt et al 2000). Such a dust lane could therefore plausibly supply a column of $\sim$ 10$^{21}$ cm$^{-2}$.

Given these three scenarios, we consider that all possibilities are plausible.

Since some of the values of R$_{l}$ (the launching radius of the wind) for different AGN in Table~\ref{filling_factors} are on similar scales to those of Scenario 3, it may be that the cold gas can exist on similar scales to the warm absorbers, this is plausible given the small filling factors of the warm absorbers.

\subsection{Conclusions}

From our {\sl XMM-Newton} RGS observations of the Seyfert 1 AGN H\,0557$-$385 we find that warm absorption is present in two phases, a low-$\xi$ phase with log $\xi$ of 0.50 and a column of $\sim$ 10$^{21}$ cm$^{-2}$ and a high-$\xi$ phase with log ionisation parameter $\xi$ of 1.62 and a column of $\sim$ 10$^{22}$ cm$^{-2}$. We detect an unresolved transition array of (M-shell) iron in both phases of the absorber. 

We infer a picture of the different scales of these components. The locations of the warm absorbers are not well constrained ($\leq$ 10$^{21}$- 3$\times10^{23}$ cm from the central continuum source), compared to the molecular torus distance of $\sim$ 10$^{18}$ cm from the source of the ionising luminosity. The Fe K$\alpha$ line and a broad H$\alpha$ line are both likely to originate at $\sim$ 10$^{15}$ cm from the black hole.


The neutral gas observed could originate from the AGN torus, the host galaxy disk, or a lane or cloud of gas above the disk.

We establish a new method of determining the volume filling factors in warm absorbers, assuming that the ionised absorbers originate as an outflow from the torus. We derive the volume filling factors for the Blustin et al (2005) sample of AGN and compare them with those published. We find that they have small volume filling factors ($\leq$ 1\%). We find reasonable agreement with the method of Blustin et al (2005). However, a group of five absorbers have filling factors derived in this paper that are much smaller than those found by B05. These absorbers all have values of log $\xi$ that are $<$ 0.7, and appear to lie out in the narrow-line region, typically hundreds of parsecs.

\section*{Acknowledgments}
\label{Acknowledgments}

CEA acknowledges financial support from a PPARC quota studentship. This work is based on observations obtained with XMM-Newton, an ESA science mission with instruments and contributions directly funded by ESA Member States and the USA (NASA). This research has made use of the NASA/IPAC Extragalactic Database (NED) which is operated by the Jet Propulsion Laboratory, California Institute of Technology, under contract with the National Aeronautics and Space Administration. This research has made use of the SIMBAD database, operated at CDS, Strasbourg, France.

\end{document}